\documentclass[11pt]{article}
\usepackage{amsfonts,latexsym}

\newfont{\blackb}{msbm10 scaled\magstep1}

\newfont{\calig}{cmsy10 scaled\magstep1}

\def\text#1{\hbox{#1}}

\newtheorem{theorem}{Theorem}[section]

\newtheorem{remark}{Remark}[section]
\newtheorem{corollary}{Corollary}[section]

\def\be{\begin{equation}}
\def\ee{\end{equation}}
\def\ben{\begin{displaymath}}
\def\een{\end{displaymath}}
\def\baa{\begin{eqnarray}}
\def\eaa{\end{eqnarray}}

\def\ba{\begin{array}}
\def\ea{\end{array}}

\def\g{\gamma}

\def\3{\ss}

\def\l{\lambda}

\def\P{\Pi}

\def\t{\tau}

\def\Th{\Theta}

\def\phi{\varphi}

\def\B{{\bf B}}
\def\C{\mathbb{C}}
\def\CP1{\mathbb{CP}^1}
\def\Z{\mathbb{Z}}

\def\P{\mathbb{P}}

\def\t0{\Theta_0}
\def\z{{\bf z}}
\def\m{{\bf m}}
\def\la{\label}

\def\f{\frac}
\def\L{{\cal L}}
\def\p{\partial}
\def\pb{{\bf p}}
\def\qb{{\bf q}}

\def\tr{{\rm tr}}
\def\0{S}
\def\1{T}

\def\log{\ln}

\def\B{{\bf B}}
\def\C{\mathbb{C}}
\def\Z{\mathbb{Z}}

\def\t0{\Theta_0}
\def\z{{\bf z}}
\def\m{{\bf m}}
\def\la{\label}

\def\f{\frac}
\def\L{{\cal L}}
\def\p{\partial}

\def\tr{{\rm tr}}
\def\0{S}
\def\1{T}

\def\log{\ln}

\def\det{{\rm det}}
\def\dbar{\bar{\partial}}
\def\CP1{\C\P 1}

\begin{document}

\begin{center}{\LARGE  Isomonodromic deformations and Hurwitz spaces}\\
\vskip1.0cm

{\large D.~A.~Korotkin}\footnote{E-mail: korotkin@discrete.concordia.ca}\\
\vskip0.5cm
Department of Mathematics and Statistics, Concordia University\\
Sherbrook West 7141, Montreal H4B 1R6 Quebec, Canada
\end{center}

\section{Introduction}

Here we solve  $N\times N$ Riemann-Hilbert (inverse monodromy) problems with
all monodromy matrices having the structure of matrices of quasi-permutation
(i.e. matrices which have only one non-zero element in each column and each row).
Such Riemann-Hilbert problem may be 
associated to arbitrary Hurwitz space of algebraic curves $\L$ of genus $g$ realized as $N$-sheeted covering over $\CP1$,
and allowes solution 
in terms of Szeg\"{o} kernel on $\L$. 
If we denote coordinate on $\CP1$ by $\l$ and projections of the branch points to complex plane by $\l_1,\dots,\l_n$ then the solution of  inverse monodromy problem of that type has the
following form:
\ben
\Psi(\l)_{jk}=S(\l^{(j)},\l_0^{(k)})E_0(\l,\l_0)\;,\hskip0.5cm j,k=1,\dots,N
\een 
where $\l^{(j)}$ is the point on $j$th sheet of $\L$ having projection $\l$ on $\CP1$;
$S(P,Q)$ is Szeg\"{o} kernel on $\L$:
\ben
S(P,Q) = \f{1}{\Th\left[^\pb_\qb\right](0)}\f{\Th\left[^\pb_\qb\right](U(P)-U(Q))}
{E(P,Q)}\;;
\een
$E(P,Q)$  ($P,Q\in\L$) is the prime-form on $\L$ and 
$E_0(\l,\l_0)=(\l-\l_0)/\sqrt{d\l d\l_0}$ is the prime-form on $\CP1$;
$\pb,\qb\in\C^g$ are two vectors such that the combination $\B\pb+\qb$
($\B$ is the matrix of $b$-periods on $\L$) does not belong to theta-divisor $(\Th)$ on Jacobi variety
$J(\L)$. 

Function $\Psi(\l)$ has determinant $1$ and is normalized at $\l=\l_0$ by the condition
$\Psi(\l=\l_0)=I$. It solves the inverse monodromy problem with quasi-permutation monodromy matrices 
which can be expressed in terms of $\pb,\qb$ and intersection indeces of certain contours on $\L$.
If parameter vectors $\pb$ and $\qb$ (and, therefore, also the monodromy matrices) don't depend on $\{\l_j\}$, we fall in the framework of isomonodromy deformations; then the residues $A_j(\{\l_j\})$ of
the function $\Psi_\l\Psi^{-1}$ satisfy the Schlesinger system.

The associate $\tau$-function can be shown to be proportional to $\Th\left[^\pb_\qb\right](0)$
up to some factor which depends only on $\{\l_j\}$. In $N=2$ case the factor can also be calculated explicitly (see \cite{KitKor})
 to give 
\be
\tau(\{\l_j\})=[\det{\cal A}]^{-\f 12}
\prod\limits_{j<k}(\l_j-\l_k)^{-\frac 18}\Theta\left[^\pb_\qb\right](0|\B)\;.
\la{tau0}\ee
where $n=2g+2$; $\l_1,\dots,\l_{2g+2}$ are branch points on the hyperelliptic curve $\L$;
${\cal A}$ is the matrix of $a$-periods of non-normalized
holomorphic differentials on this curve.

 As it was demonstrated by Malgrange \cite{Malg80}, the tau-function
of Schlesinger system may be interpreted as determinant of certain Toeplitz operator. It was further argued by Palmer \cite{Palm90}
that the tau-function could also be interpreted as determinant of Cauchy-Riemann operator acting on certain class
of matrix spinors with  prescribed singularities at certain points on Riemann sphere. 
However, the non-standard type of the domain of the Cauchy-Riemann operators defined in this way makes it rather difficult  to establish the links
with more conventional framework  of \cite{Quillen}.

On the other hand, the Cauchy-Riemann determinants corresponding to compact Riemann surfaces were very actively exploited 
in  the context of perturbartive string theory in late 80's (see \cite{Knizhnik,Alvares,BeiShk,Fay2}.
Comparison with formulas of works \cite{Knizhnik,Alvares,BeiShk} shows that formula (\ref{tau0}) coincides with 
 the determinant of  Cauchy-Riemann operator acting on  on spinors on $\L$ which have
twists $e^{-2\pi i p_j}$ and $e^{2\pi i q_j}$ along cycles $a_j$ and $b_j$  respectively. 
Therefore, it seems tempting to speculate that this observation is also true for arbitrary curves;
this should be a subject of further study.

Another result of these notes concerns the divisor $(\vartheta)\subset \C^n$ in the
space of parameters $\{\l_j\}$ introduced by Malgrange. This is the divisor of zeros of $\tau$-function
in $\C^n$, or, equivalently, divisor in $\{\l_j\}$-space
 where the solution of inverse monodromy problem with given monodromy data fails to exist.
For our class of monodromy data we have
\ben
\{\l_j\}\in (\vartheta) \;\;  \Leftrightarrow \;\; \B\pb+\qb \in (\Th)\;,
\een
where $(\Th)$ is theta-divisor on Jacobian $J(\L)$.

\section{Schlesinger system and $\tau$-function}

Let us fix the notations. Consider the following Riemann-Hilbert problem on $\CP1$: for a given set of 
$n+1$ points $\l_0,\l_1,\dots,\l_n\in\C$, construct a function $\Psi(\l): \CP1\setminus\{\l_1,\dots,\l_n\}\to
SL(N,\C)$, which has the following properties:
\newline
- $\Psi(\l)$ is holomorphic on universal covering of
 $\l\in\CP1\setminus\{\l_1,\dots,\l_n\}$ and on some sheet of this covering $\Psi(\l_0)=I$.
\newline
- $\Psi(\l)$ has regular singular points at $\l=\l_j$, $j=1,\dots,n$ with given monodromy matrices 
$M_j\in SL(N,\C)$.

If in addition to monodromy matrices we fix the logarithms of their eigenvalues, this RH problem is 
always solvable outside of submanifold of codimension 1 in the space of parameters $\{\l_j,M_j\}$.
Outside of this submanifold function $\Psi$ satisfies the matrix differential equation
\be
\f{d\Psi}{d\l}=\sum_{j=1}^n\left(\f{A_j}{\l-\l_j}-\f{A_j}{\l_0-\l_j}\right)\Psi
\la{ODE}\ee
with certain matrices $A_j\in sl(N,\C)$; eigenvalues $t_j^{(1)},\dots,t_j^{(N)}$
 of $A_j$ are equal (up to the factor $2\pi i$) to the
logarithms of eigenvalues of matrices $M_j$. We call the set $\{M_j,t_j^{(k)}\}$ the 
monodromy data.

If we impose the isomonodromy conditions, ${d M_k}/{d\l_j}=0\;$, $j,k=1,\dots,n$, and assume that
$t_j^{(l)}-t_j^{(s)}\not\in \Z$ for any $l$ and $s$, then function $\Psi$ satisfies the following
equations with respect to $\l_j$:
\be
\f{d\Psi}{d\l_j}=\left(\f{A_j}{\l_0-\l_j}-\f{A_j}{\l-\l_j}\right)\Psi
\la{ODE1}\ee
Compatibility condition of (\ref{ODE}) and (\ref{ODE1}) gives Schlesinger system
for the residues $A_j$ as functions of poles $\{\l_k\}$.

In particular, if we choose $\l_0=\infty$, the Schlesinger system has the following form:
\ben
\f{\p A_j}{\p\l_k}=\f{[A_j, A_k]}{\l_j-\l_k}\;,\hskip0.6cm j\neq k\;;\hskip0.8cm
\f{\p A_j}{\p\l_j}=-\sum_{k\neq j} \f{[A_j, A_k]}{\l_j-\l_k}\;.
\een

The $\tau$-function of Schlesinger system is defined by the formula \cite{JMS}:
\be
\f{d}{d\l_j}\log\tau=\f{1}{2} {\rm res}\Big|_{\l=\l_j}\tr\left(\Psi_{\l}\Psi^{-1}\right)^2\equiv
\sum_{j<k}\f{\tr A_j A_k}{\l_j-\l_k}\;.
\la{deftau}\ee
According to Malgrange \cite{Malg80}, the function $\tau(\{\l_j\})$ vanishes in the space 
$\C^n\setminus\{\l_j=\l_k\;,\; j,k=1,\dots,n\}$ precisely on the submanifold where function $\Psi$
corresponding to a given set of monodromies $M_j$ and eigenvalues $t_j^{(k)}$ fails to exist.

\section{Riemann-Hilbert problems associated to hyperelliptic curves}

In this section we give a modified version of construction proposed in   \cite{KitKor}. 
Take $n=2g+2$ and consider hyperelliptic curve $\L$ given by equation
\be
w^2=\prod_{j=1}^{2g+2}(\l-\l_j)\;.
\la{hyper}\ee
Let us define two $-1/2\;$-forms  $\phi_{1,2}$  in fundamental polygon $\hat{\L}$ of $\L$ by the 
formulas:
\baa
\phi_{1}(P)=\Th\left[^\pb_\qb\right]\left(U(P) + U(D_1)\right)E(P,D_1)\\
\phi_{2}(P)=\Th\left[^\pb_\qb\right]\left(U(P) + U(D_2)\right)E(P,D_2)
\la{phi1}
\eaa
where $\pb,\qb\in\C^g$; $D_1$ and $D_2$ are two arbitrary points of curve $\L$; $E(P,Q)$ is the prime form on $\L$;
initial point of the Abel map $U(P)$ is chosen to be $\l_1$.
Define auxiliary $2\times 2$ function $\Phi(\l)$:
$$
\Phi_{kj}(\l)=\phi_k(\l^{(j)})\;,
$$
where $k,j=1,2$; $\l^{(j)}$ denotes point of $\L$ belonging to $j$th sheet and having projection $\l$ 
on $\CP1$. Define function  $\Psi(\l)$ by the formula
\be
\Psi(\l)=\sqrt{\frac{\det\Phi(\infty^1)}{\det\Phi(\l)}}\Phi^{-1}(\infty^1)\Phi(\l)\;.
\la{Psi2}\ee
The $1/2$-differentials in the denominator of prime-form in $\phi_{1,2}$ cancel out in expression for $\Psi$; thus  $\Psi$  is a function i.e. 0-form on $\L$.
The following theorem takes place, which is slightly modified version of the statement formulated in \cite{KitKor}\footnote{In a different form solution of the same RH problem was obtained in \cite{DIKZ}.}.
\begin{theorem}
Let us fix some points $\l_1,\dots,\l_{2g+2}\in\C$ and vectors $\pb,\qb\in\C^g$. 
Consider hyperelliptic curve $\L$  (\ref{hyper}) with   matrix of $b$-periods $\B$.
Assume that $\Th\left[^\pb_\qb\right](0|\B)\neq 0$ i.e. vector $\B\pb+\qb$ does not belong to theta-divisor.
Then function $\Psi(\l)$ defined by (\ref{phi1}),(\ref{Psi2}) gives a solution to matrix Riemann-Hilbert
problem with $\l_0=\infty$ and singularities at the points $\l_1,\dots,\l_{2g+2}$ with off-diagonal monodromies 
\be
M_j= \left(\ba{cc} 0 & -m_j \\m_j^{-1} & 0 \ea\right)\;,
\la{Mj1}
\ee
where
\ben
m_1=1,\;\;\;\;\;\;\;
m_2=\exp\{-2\pi i \sum_{k=1}^g p_k\},
\een
\ben
m_{2j+1}= - \exp\{2\pi i q_j- 2\pi i \sum_{k=j}^g p_k\},
\een
\be
m_{2j+2}= \exp\{2\pi i q_j-2\pi i \sum_{k=j+1}^g p_k\},
\la{mj}\ee
\end{theorem}
{\it Proof.} 
We can rewrite the expression for  ${\rm det}\Phi$  using Fay identities \cite{Fay}:
\ben
\Th(\z+U(c)-U(a))\Th(\z+U(d)-U(c))E(c,b)E(a,d)
\een
\ben
+\Th(\z+U(c)-U(b))\Th(\z+U(d)-U(a))E(c,a)E(d,b)
\een
\ben
=\Th(\z+U(c)+U(d)-U(a)-U(b))\Th(\z)E(c,d)E(a,b)\;.
\een
where $\z\in \C^g$; $a,b,c,d$ are four arbitrary points of $\L$.
After identification $-\z\equiv\B\pb+\qb$, $a\equiv D_1$, $b\equiv D_2$, $c\equiv P$, $d\equiv P^*$, 
the left-hand side of Fay identities gives $\det\Phi(P)$. Evaluating the right-hand side we obtain
\be
{\rm det} \Phi(P)= \Th\left[^\pb_\qb\right](0)\Th\left[^\pb_\qb\right](U(D_1)+U(D_2))E(P,P^*)E(D_1,D_2)\;.
\la{dethyp}\ee
Since function $\Psi$ is independent of $D_1$ and $D_2$, function $\Psi$ is undefined
precisely at the
points where the first prefactor vanishes i.e. vector $\B\pb+\qb$ belongs to the theta-divisor $(\Th)$
on Jacobian $J(\L)$. Outside of this singular variety function $\Psi$ is well-defined, non-singular
and invertible in $\l$-plane outside of the points $\l_j$.  At the points $\l_j$ it has regular
singularities; expressions for monodromy matrices (\ref{mj})  follow from periodicity properties of
theta-function.

If we assume that vectors $\pb$ and $\qb$ are $\{\l_j\}$-independent, functions
$A_j(\{\l_k\})\equiv {\rm res}|_{\l=\l_j} \Psi_\l\Psi^{-1}$ satisfy the Schlesinger system;
corresponding tau-function is given by the following theorem.

\begin{theorem}\cite{KitKor}
The tau-function of Schlesinger system corresponding to monodromy matrices (\ref{mj}) is given by
\be
\tau(\{\g_j\})=[\det{\cal A}]^{-\f 12}
\prod\limits_{j<k}(\l_j-\l_k)^{-\frac 18}\Theta\left[^\pb_\qb\right](0|\B)\;.
\la{tau}\ee
i.e. coincides with determinant of Cauchy-Riemann operators $\partial_{1/2}^{\pb,\qb}$ on $\L$ \cite{Knizhnik,Alvares,BeiShk}.
\end{theorem}
{\it Proof}. Here we give a version of the proof which is slightly simplified comparing with the original version of (\cite{KitKor}).
Taking into account the following identity valid for $2\times 2$ matrices,
\ben
\frac{1}{2}\tr(\Psi_{\l}\Psi^{-1})^2 = -\frac{\det(\Phi_\l)}{\det\Phi}+\frac{1}{4}\left(\frac{(\det\Phi)_\l}{\det\Phi}\right)^2\;,
\een
we find that
\be
\frac{1}{2}\tr(\Psi_{\l}\Psi^{-1})^2(\l) 
= -\frac{1}{\Th\left[^\pb_\qb\right](0)}\frac{\partial^2\left\{\Th\left[^\pb_\qb\right](U(\mu)-U(\l))\right\}}{\partial\l\partial\mu}
\Big|_{\mu=\l}-\f{\partial^2\left\{\log E(\l,\mu)\right\}}{\partial\l\partial\mu}\Big|_{\mu=\l^*}
\la{trace}\ee
Dependence of $\tau$-function on vectors $\pb$ and $\qb$ is contained in the first term of right-hand side. This term can be further rewritten as
\ben
\f{1}{\Th\left[^\pb_\qb\right](0)}\sum_{k,l=1}^g\frac{\partial^2\Th\left[^\pb_\qb\right](0)}{\p z_k\p z_l}\f{d U_k}{d\l}\f{d U_l}{d\l}\equiv
4\pi i\sum_{k,l=1}^g\frac{\partial\log\Th\left[^\pb_\qb\right](0)}{\p\B_{lk}}\f{d U_k}{d\l}\f{d U_l}{d\l}
\;,
\een
where we used the heat equation for theta-function; $z_k$ denotes the $k$th argument of theta-function. Dependence of matrix of $b$-periods on the branch points is given by the following equations \cite{SpeSch,KitKor}:
\ben
\f{\partial\B_{kl}}{\p\l_j}=\pi i \frac{\p U_k}{\p\kappa_j}(\l_j)\frac{\p U_l}{\p\kappa_j}(\l_j)\;,
\een
where $\kappa_j=\sqrt{\l-\l_j}$ is a local parameter near point $\l_j$. On the other hand, value
$
4\frac{\p U_k}{\p\kappa_j}(\l_j)\frac{\p U_l}{\p\kappa_j}(\l_j)
$
is nothing but the residue of the rational function 
$
\f{d U_k}{d\l}(\l)\f{d U_l}{d\l}(\l)
$
at $\l=\l_j$.
Continuing the calculation of the first term in (\ref{trace}) we get
\ben
\sum_{j=1}^g \f{1}{\l-\l_j}\f{\p\log\Theta\left[^\pb_\qb\right](0)}{\p\l_j}\;,
\een
and, therefore,
\be
\tau=f(\{\l_j\})\Theta\left[^\pb_\qb\right](0)\;,
\la{tau1}\ee
where function $f$ does not carry any dependence on $\pb$ and $\qb$.
Now, to determine function $f$ we can choose vectors $\pb$ and $\qb$ in such a way that the tau-function may be explicitly calculated in elementary functions.
One of possible choices of that kind is to take $\pb,\qb$ to coincide with some even half-integer characteristic $\pb^T, \qb^T$. We choose characteristic $\pb_T, \qb_T$ to correspond to some subset $T=\{i_1,\dots,i_{g+1}\}$ of
the set $\{1,\dots,2g+2\}$ via the standard relation
\ben
\B\pb^T +\qb^T = U(\l_{i_1})+\dots+U(\l_{i_{g+1}})-K\;.
\een
According to Thomae formulas \cite{Fay},
\ben
\Th^4\left[^{\pb^T}_{\qb^T}\right](0)=\pm\frac{(\det{\cal A})^2}{(2\pi i)^{2g}}\prod_{j,k\in T}
(\l_j-\l_k)\prod_{j,k\not\in T}(\l_j-\l_k)\;,
\een
where ${\cal A}_{jk}=\oint_{a_k}\f{\l^{k-1} d\l}{w}$.
Therefore, the $\tau$-function (\ref{tau1}) may be up to unessential overall constant factor
rewritten as follows:
\be
\tau=f(\{\l_j\})(\det{\cal A})^{1/2}\prod_{\l_j,\l_k\in T}
(\l_j-\l_k)^{1/4}\prod_{\l_j,\l_k\not\in T}(\l_j-\l_k)^{1/4}\;.
\la{tau2}\ee
Alternatively,  we can easily calculate the same $\tau$-function directly.
Taking into account that $U(\l_1)=0\;\;$; $U(\l_2)=\frac{1}{2}\sum_{k=1}^g {\bf e}_k\;\;$;
$U(\l_{2j+1})=\f{1}{2}\B{\bf e}_j+\f{1}{2}\sum_{k=j}^g {\bf e}_j\;\;$; 
$U(\l_{2j+2})=\f{1}{2}\B{\bf e}_j+\f{1}{2}\sum_{k=j+1}^g {\bf e}_j$, we find:
\ben
q^T_{j+1}- q_j^T =\frac{1}{2}(\delta_{2j+2}+\delta_{2j+3}+1)\;;\hskip0.7cm
p_j^T=\frac{1}{2}(\delta_{2j+1}+\delta_{2j+2}+1)\;,
\een
where $\delta_j=1$ for $j\in T$ and $\delta_j=0$ for $j\not\in T$. 
Substituting these formulas to (\ref{mj}) we see that the 
monodromy matrices have the following form:
\ben
M_j= i(-1)^{\delta_j+\delta_1}\sigma_1\;,
\een
where by $\sigma_j$, $j=1,2,3$ we
denote  the standard Pauli matrices. 
By simultaneous similarity transformation which does not modify associate $\tau$-function this set
of monodromy matrices may be transformed to the set of diagonal matrices
\ben
\tilde{M}_j = i\sigma_3\;,\hskip0.6cm \l_j\in T\;;\hskip0.6cm
\tilde{M}_j = -i\sigma_3\;,\hskip0.6cm \l_j\not\in T\;.
\een
The associate function $\Psi$ may be chosen to be diagonal: 
$\Psi(\l)={\rm diag}(\phi_0(\l),\phi_0^{-1}(\l))$ with
\ben
\phi_0(\l)=\prod_{j\in T}(\l-\l_j)^{1/4}
\prod_{j\not\in T}(\l-\l_j)^{-1/4}\;,
\een
which leads to the following formula for $\tau$-function:
\be
\tau=\prod_{j,k\in T}(\l_j-\l_k)^{1/8}
\prod_{j,k\not\in T}(\l_j-\l_k)^{1/8}
\prod_{j\in T,\;k\not\in T}(\l_j-\l_k)^{-1/8}\;.
\la{tau3}\ee
Comparing (\ref{tau2}) and (\ref{tau3}) we get
\ben
f(\{\l_j\})= (\det {\cal A})^{-1/2}\prod_{j<k}(\l_j-\l_k)^{-1/8}\;,
\een
proving (\ref{tau}).

\section{Solution of matrix Riemann-Hilbert problems in terms of Szeg\"{o} kernel on algebraic curves}

Consider non-singular algebraic curve $\L$ 
defined by polynomial equation
\ben
f(\l,w)=0
\een
of degree $N$ in $w$. Hurwitz space is the moduli space of curves of fixed genus $g$ and fixed number of sheets $N$. In analogy to Dubrovin \cite{Dub94} we shall in addition fix the types of ramification 
at all branch points. Denote projections of branch points on $\l$-plane by  $\l_j$, $j=1,\dots,n$
(admitting little inaccuracy we shall also call $\l_j$ the branch points). 
If we denote multiplicities of branch points $\l_1,\dots,\l_n$ by
$m_1,\dots,m_n$ respectively, the genus of $\L$ is given by Riemann-Hurwitz formula
\ben
g=\frac{1}{2}\sum_{j=1}^n m_j - N + 1\;.
\een
We consider Hurwitz space $H$ consisting of the curves which can be obtained from $\L$ by variation of branch points $\l_j$ without changing their type of ramification.
Assume that the normalization point $\l_0$ does not coincide with any of $\l_j$.

To each Hurwitz space $H$ we can associate solution of certain RH problem with singularities at points
$\l_j$ and quasi-permutation monodromy matrices 
(For brevity we call any matrix which has only one non-vanishing element in each row and only one 
non-vanishing element in each column the quasi-permutation matrix.)

Denote by $\pi\;:\;\L\to\CP1$ the  projection of $\L$ to $\l$-plane.
Let us denote by $l_1,\dots,l_n$ the natural basis in $H^0(\CP1\setminus \{\l_1,\dots,\l_n\},\Z)$.
As a starting point of all $l_j$ we choose $\l_0$. Consider $\pi^{-1}(l_j)$. This is a set 
of $N$ non-intersecting contours $l_j^{(k)}$, $k=1,\dots,N$ on $\L$, where by $l_j^{(k)}$ we
denote contour starting at $\l_0^{(k)}$. Denote the endpoint of  $l_j^{(k)}$ by $\l_0^{(k')}$ with
some $k'=k'(k)$.
If $\l_j^{(k)}$ is not a branch point, then $k=k'$, and contour $l_j^{(k)}$ is closed; if $\l_j^{(k)}$ is a branch point, then $k\neq k'$ and contour $l_j^{(k)}$ is non-closed.

Assume now that point $\l_0$ does not belong to the set of projections of basic cycles $(a_j,b_j)$ on
$\CP1$. Introduce intersection indexes
\baa
\alpha_{js}^{(k)}=l_j^{(k)}\circ a_s\;,\hskip0.6cm
\beta_{js}^{(k)}=l_j^{(k)}\circ b_s\\{\rm where}\;\;\;
j=1,\dots,n\;;\;\;s=1,\dots,g\;;\;\;k=1,\dots,N
\la{inter}\eaa

Choose on $\L$ a canonical basis of cycles $(a_j,b_j),\;j=1,\dots,g$. Introduce the    basis of holomorphic 1-forms $dU_j$ on $\L$
normalized by $\oint_{a_j}dU_k=\delta_{jk}$, matrix of $b$-periods $\B$ and the Abel map $U(P)\,,\;P\in\L$. Denote initial point of Abel map by $P_0$.

Let us introduce function $\Psi(\l)$ in analogy to (\ref{Psi2}):
\be
\Psi(\l)=\left[\f{\det\Phi(\l_0)}{\det\Phi(\l)}\right]^{1/N}\Phi^{-1}(\l_0)\Phi(\l)\;.
\la{PsiN}\ee
Function $\Phi(\l)$ is defined as follows:
\be
\Phi(\l)_{kj}\equiv \frac{\l-\mu}{\sqrt{d\l d\mu}}\phi_k (\l^{(j)})\; ,
\ee
 where by $\l^{(j)}$ we denote the point of $j$th sheet of curve $\L$ having projection $\l$ on $\CP1$;
$\mu\in\C$ is an arbitrary point.
Here $\phi_k(P)$  are holomorphic spinors on $\L$. To define them choose an arbitrary set of $N$ positive non-special divisors $D_k$, $k=1,\dots,N$ of degree $N-1$ each i.e 
$D_k=\sum_{j=1}^{N-1} D_k^j$. Take
\be
\phi_k(P)=
\frac{\Th\left[^\pb_\qb\right](U(P)+U(D_k)-C)\prod_{j=1}^{N-1} E(P, D_k^j)}
{\prod_{j=1}^N E(P,\mu^{(j)})}
\la{phik}
\ee
where 
$\pb,\qb\in \C^g$;  $E(P,Q)$ is the prime-form;
$
C\equiv\sum_{k=1}^N U(\l^{(k)})\,.
$

It is clear that vector $C$ does not depend on $\l$; it depends only on the choice of initial point of Abel map $P_0$. This follows from the fact that for any holomorphic 1-form $dU(P)$ on $\L$ the sum
$\sum_{j=1}^N dU(\l^{(j)})$ is holomorphic 1-form on $\CP1$, therefore identically vanishing.
The function $\phi_k(\l^{(j)})/\sqrt{d\l}$ behaves near branch point $\l_j$ as $\tau_j^{-m_j/2}$
where $\tau_j=(\l-\l_j)^{1/(m_j+1)}$ is the local parameter near $\l_j$. Therefore, to completely
define this function on $\hat{\L}$, one has to define system of contours on $\L$ which connect the
branch points with odd $m_j$, and where functions  $\phi_k(P)/\sqrt{d\l}$ change sign. Denote this system
of contours by $L$.

\begin{theorem}
Suppose that $\Th\left[^\pb_\qb\right](0)\neq 0$. Then
function $\Psi$ (\ref{PsiN}) is independent of the choice of divisors $D_k$ and point $\mu$ and solves the RH problem on $\CP1$ with  quasi-permutation matrices $M_j$ which can be expressed in terms of
vectors $\pb$ and $\qb$.
\end{theorem}
{\it Proof.} By counting number of poles and zeros it is easy to check that $\det \Phi(\l)$ does not vanish
outside of branch points $\l_j$. 
The spinors $\phi_k(P)$ (\ref{phik}) transform as follows under the analytical continuation along
basic cycles:
\be
T_{a_j}[\phi_k(P)]= e^{2\pi i p_j}\phi_k(P)\;,\hskip0.6cm
T_{b_j}[\phi_k(P)]= e^{-2\pi i q_j}\phi_k(P)\;.
\la{Tphi}\ee
When we consider analytical continuation of $\psi(\l^{(k)})/\sqrt{d\l}$ along contour $l_j^{(k)}$ from
$\l_0^{(k)}$ to $\l_0^{(k')}$, we come to the value $\psi(\l_0^{(k')})/\sqrt{d\l}$ up to the factor
which is collected from crossing the contours $\{a_j,b_j\}$ and contour $L$, where this function has jumps.
Denote by $I_j^{(k)}$ the intersection index of $l_j^{(k)}$ and $L$.
Then the total factor  we collect along contour $\l_j^{(k)}$ is $\exp\left\{\pi i I_j^{(k)}+2\pi i \left[\sum_{s=1}^g \alpha_{js}^{(k)} q_s+
\beta_{js}^{(k)} p_s\right]\right\}$, where intersection indeces $\alpha_{js}^{(k)}$ and $\beta_{js}^{(k)}$ are given by (\ref{inter}). Therefore, monodromy matrices 
corresponding to our $\Psi$, have the following form
\be
(M_j)_{kl}=\exp\left\{\pi i I_j^{(k)}+2\pi i \left[\sum_{s=1}^g \alpha_{js}^{(k)} q_s+
\beta_{js}^{(k)} p_s\right]\right\}\delta_{\tilde{k}(k),l}
\la{Mjpq}\ee
($\delta_{ab}$ is the Kronecker symbol); obviously, this is a matrix of quasi-permutation.
Independence of function $\Psi$ on the choice of divisors $D_k$ and point $\mu$ follows from uniqueness
of solution of Riemann-Hilbert problem with given $\{M_j\}$ and $\{t^{(s)}_j\}$. 

Condition $\Th\left[^\pb_\qb\right](0)\neq 0$ of the theorem guarantees the non-vanishing of 
$\det\Phi(\l)$ in (\ref{PsiN}). Namely, for arbitrary $N$ points $P_j\in\L$ we can prove that
\ben
\det_{N\times N}\{\Th\left[^\pb_\qb\right](U(P_j)+U(D_k)-C)\prod_{j=1}^{N-1} E(P_j, D_k^j)\}
\een
\be
= F(\mu,\{\l_j\},\{D_k\})\;
\Th\left[^\pb_\qb\right](\sum_{j=1}^N U(P_j)-C)\prod_{j,k=1}^{N} E(P_j, P_k)
\la{det1}
\ee
for some $\{P_j\}$-independent section $F$. The proof of formula (\ref{det1}) may be obtained in a 
standard way. First, it is easy to prove that the r.h.s. and l.h.s. are sections of the same bundle
on $\L$ with respect to each $P_j$. Then we check that positions of zeros of l.h.s. and r.h.s.
with respect to each $P_j$ coincide.
Choosing $P_j=\l^{(j)}$ we get $\sum_{j=1}^N U(P_j)=C$; therefore, $\det\Phi(\l)$ is proportional to $\Th\left[^\pb_\qb\right](0)$ as in $2\times 2$ case (\ref{dethyp}). Thus function $\Psi$ (\ref{PsiN}) 
is undefined if $\Th\left[^\pb_\qb\right](0)=0$ i.e. 
\ben
\B\pb+\qb\in (\Th)
\een
where $(\Th)$ is theta-divisor on Jacobian of $\L$. 
\vskip0.5cm
The previous construction of function $\Psi$ may be simplified by choosing
$\mu=\l_0$, and $D_k=\sum_{j\neq k} \l_0^{(j)}$. 
\begin{corollary}
Suppose that $\Th\left[^\pb_\qb\right](0)\neq 0$.
Then function $\Psi(\l)$ with components
\be
\Psi(\l)_{kj}=\f{1}{\Th\left[^\pb_\qb\right](0)}\f{\Th\left[^\pb_\qb\right](U(\l^{(j)})-U(\l_0^{(k)}))}
{E(\l^{(j)},\l_0^{(k)})}\frac{\l-\l_0}{\sqrt{d\l d\l_0}}
\la{psikj}
\ee
belongs to $SL(N,\C)$ for any $\l\in\C$, is non-singular on $\C$ outside of points $\l=\l_j$,
satisfies normalization condition $\Psi(\l_0)=I$ and solves Riemann-Hilbert problem with monodromy matrices (\ref{Mjpq}).
\end{corollary} 
\begin{remark}\rm
Formula (\ref{psikj}) may be  rewritten in terms of Szeg\"{o} kernel on $\L$:
\be
S(P,Q) = \f{1}{\Th\left[^\pb_\qb\right](0)}\f{\Th\left[^\pb_\qb\right](U(P)-U(Q))}{E(P,Q)}\;,
\la{szego}\ee
which is $(1/2,1/2)$ differential on $\L\times\L$, as follows:
\be
\Psi(\l)_{kj} = S(\l^{(j)},\l_0^{(k)})E_0(\l,\l_0)
\la{psisz}
\ee
where $E_0(\l,\l_0)=(\l-\l_0)/{\sqrt{d\l d\l_0}}$ is the prime-form on $\CP1$.
\end{remark}
{\it Proof of the Corollary.}
For any two sets $P_1,\dots,P_N$ and $Q_1,\dots,Q_N$ 
we have the following identity (see \cite{Fay}, p.33):
\be
\det\{S(P_j,Q_k)\}= \frac{\Th\left[^\pb_\qb\right]\left(\sum_{j=1}^N (U(P_j)-U(Q_j))\right)}
{\Th\left[^\pb_\qb\right](0)}\frac{\prod_{j<k} E(P_j,P_k) E(Q_k,Q_j)}{\prod_{j,k} E(P_j,Q_k)}
\la{ident}\ee
analogous to(\ref{det1}).
Choosing $P_j\equiv \l^{(j)}$ and $Q_k\equiv \l_0^{(k)}$ and using the basic properties of prime-form
we conclude that $\det\Psi(\l)=1$. 
Normalization condition $\psi_j\left(\l_0^{(k)}\right)=\delta_{jk}$ is the corollary 
 of asymptotic expansion of prime form:
\ben
E(P,Q)=\frac{z(P)-z(Q)}{\sqrt{dz(P) dz(Q)}}(1+ o(1))
\een
as $P\to Q$, where $z(P)$ is a local parameter.

Let us consider separately the case $\l_0=\infty$. In this case the above formulas should be slightly modified.
\begin{corollary}
Suppose that $\Th\left[^\pb_\qb\right](0)\neq 0$.
Define function $\Psi(\l)$ with components
\be
\Psi(\l)_{kj}=\f{1}{\Th\left[^\pb_\qb\right](0)}\f{\Th\left[^\pb_\qb\right](U(\l^{(j)})-U(\infty^{(k)})}
{\Th[S](U(\l^{(j)})-U(\infty^{(k)})}\sqrt{\f{d W(\l^{(k)})}{d(1/\l)}(\infty)}
\sqrt{-\f{d W(\l^{(j)})}{d\l}}
\la{psiinf}
\ee
where $[S]$ is an arbitrary non-degenerate odd half-integer characteristic and 
$dW(P)=\sum_{j=1}^g\frac{\p\Th[S]}{\p z_j}(0) dU_j$. Then function $\Psi(\l)$
belongs to $SL(N,\C)$ for any $\l\in\C$, is non-singular on $\C$ outside of points $\l=\l_j$,
satisfies normalization condition $\Psi(\infty)=I$ and solves Riemann-Hilbert problem with monodromy matrices (\ref{Mjpq}).
\end{corollary} 

If we now assume that vectors $\pb$ and $\qb$ don't depend on $\{\l_j\}$, matrices $M_j$ also
don't carry any $\{\l_j\}$-dependence and the   isomonodromy deformation equations take place.

\begin{theorem}
Assume that vectors $\pb$ and $\qb$ don't depend on $\{\l_j\}$. Then
functions 
\be
A_j(\{\l_j\}) \equiv {\rm res}|_{\l=\l_j} \{\Psi_{\l}\Psi^{-1}\}
\la{solA}\ee
where function $\Psi(\l)$ is defined in (\ref{psiinf}), 
satisfy Schlesinger system 
outside of hyperplanes $\l_k=\l_j$ and submanifold of codimension one,
 on which vector $\B\pb+\qb$ belongs to theta-divisor $(\Th)$ on $\L$.
\end{theorem}

Let us discuss now the calculation of corresponding $\tau$-function.
It is known \cite{Malg80} that the $\tau$-function vanishes outside of the hyperplanes
$\l_j=\l_k$ precisely at those points where the Riemann-Hilbert problem does not have a solution;
together with explicit calculations in $2\times 2$ case this suggests that the tau-function should be
proportional to $\Th\left[^\pb_\qb\right](0)$. Explicit calculation shows that this is really the case,
and, moreover, this factor contains the whole dependence of $\tau$ on vectors $\pb$ and $\qb$. 
So, 
\ben
\tau= f(\{\l_j\}) \Th\left[^\pb_\qb\right](0)
\een
with some function $f$ depending only on $\{\l_j\}$. Explicit calculation of function $f$ is 
possible in
some special cases, like the curves of $\Z_N$ class \cite{BerRad88}.

Taking into account the coincidence of the $\tau$-function in $2\times 2$ case with determinant of Cauchy-Riemann operator 
acting on $1/2$-forms $w(P)$ on $\L$ satisfying boundary conditions $w(P+a_j) = e^{2\pi i p_j}w(P)$, 
$w(P+b_j) = e^{-2\pi i q_j}w(P)$ i.e.
\be
\tau=\det \dbar_{1/2}^{\pb,\qb} 
\ee
it is tempting to suggest that this coincidence takes place for arbitrary curves; then 
function $f$ would coincide (see \cite{Knizhnik,Alvares,BeiShk}) with $[\det\dbar_0]^{-1/2}$
where operator $\dbar_0$ acts on 0-forms on $\L$. 
\begin{remark}
It is clear that $\det \dbar_{1/2}^{\pb,\qb}$, as well as $\tau$-function, vanishes 
if $\B\pb+\qb\in (\Th)$, since in this case
$1/2$-form $\Th\left[^\pb_\qb\right](U(P)-U(Q))/E(P,Q)$ for any $Q\in\L$ belongs to its kernel.
\end{remark}
Another argument suggesting possible coincidence of $\tau$ and $\det \dbar_{1/2}^{\pb,\qb}$ in general case arises from consideration of Palmer \cite{Palm90}.
It is also relevant to notice that close link between Cauchy-Riemann determinants and tau-functions
arising in the theory of KP equation was mentioned in \cite{GriOrl}. 

Finally, following \cite{Malg80}, denote the divisor of zeros of $\tau$-function in
$\C^n$ by $(\vartheta)$. Then we get the following relationship
between Malgrange's divisor $(\vartheta)$ and theta-divisor $(\Th)$ on Jacobian $J(\L)$:
\ben
\{\l_j\}\in (\vartheta) \;\;  \Leftrightarrow \;\; \B\pb+\qb \in (\Th)\;,
\een

{\bf Acknowledgements} I thank John Harnad, Alexey Kokotov and Alexandr Orlov for important discussions at different stages of this work.

\end{document}